\documentclass[letterpaper,10pt]{article}

\usepackage{geometry}
\usepackage[numbers]{natbib}
\usepackage{amsmath,amsfonts,amssymb}
\usepackage{theorem}

\usepackage{graphicx}

\newtheorem{Theorem}{Theorem}[section]
\newtheorem{Proposition}{Proposition}[section]
\newtheorem{Lemma}{Lemma}[section]

\numberwithin{equation}{section}

\title{\bfseries Optimal tests in $AR(m)$ time series model}
\author{  Tewfik LOUNIS \\ LMNO. CNRS UMR  6139 Universit\'e de Caen. }

\begin{document}

\maketitle

\begin{abstract}
\noindent A method for an evaluation of the error between an unknown
parameter and its estimator is developed. Its application enables
us to preserve  the asymptotic power of a constructed  test.
Testing problems in $AR(1)$ and $ARCH$ models are studied with a
derivation of the asymptotic power function. Also the results are
extended to $AR(m)$ time series model.\\

\end{abstract}

\section*{Introduction}

\indent A great deal of data in economic, biological, financial,
hydrological, biomedical occurs in form of time series which takes
into account several criteria, such as, for instance, the
dependance of the observations. Several works search were devoted
for the problem of estimating of the unknown parameter of a time
series model, namely, we cite the classical one such as the
maximum likelihood estimates, the Yule-Walker estimates, the
M-estimates and the least square estimates. In statistical, in the
study of testing problem, a large variety of tests use the cited
classical estimators. In the local asymptotic normality (Lan) of
log likelihood ratio, central sequence gives the expression of the
constructed test and its power function. Often the estimated
central sequence depends on the cited estimators. The considering
of the time  series model is followed by the constructing and the
study of its asymptotic proprieties. For a testing procedure, one
of the desirable and important propriety is the asymptotic
optimality. About the local asymptotic normality the interested
reader may refer to \cite{l}, \cite{Swensen1985}, \cite{HM} and
references therein. Also the derivation of the most important
proprieties of statistic tests is obtained when the parameter of
the time series model is known. In a general case, this parameter
remains unspecified. Its estimation induces an error. This latter
alters the asymptotic power of the statistic test. A good
evaluation of this error enables us to avoid this
effect.\\
 \indent The objective of this paper is to develop a method for a good
 evaluation of the error  between an unspecified parameter and its
 estimator.  Based on the confidence intervals
and the simultaneous confidence intervals, the evaluation will be
established for the univariate and multivariate parameters respectively.\\
 This work presents several applications of our main results,
such us the equivalence between the central and the estimated
central sequences which appears in the log-likelihood ratio. Thus,
enables as to preserve asymptotically the power function of a
constructed test. A great importance was devoted for the choice of
the version of the Lan. More precisely, for the univariate case,
we use the results of \cite{hb} and for the multivariate case, we
refer to \cite{fl}. We focus our study on the $AR(1)$ time series
model with an extension to $ARCH$ models. In a last, a
generalization will be obtained for the $AR(m)$ time series
models. In each case, the optimality is asserted, and the
asymptotic power function is derived.
\\ The paper is structured as
follows. In Section \ref{evaluation univa}, on a basis on the
confidence intervals, we develop a method for the evaluation of
the error between  an unspecified parameter and its estimator. In
Section \ref{testuniv}, we apply our method in testing problem.
Section \ref{general} is devoted for the generalization of our
results. In Section \ref{simul}, simulations are carried out to
aim to investigate the performance of our methodology. All
mathematical developments are relegated to the Section
\ref{demonstration} .
\section{Evaluation of the error for the univariate parameter
}\label{evaluation univa}

\indent Let $Y_1,\dots,Y_n,\dots$ be a stochastic model with a
distribution $P=P_\theta$ that depends on a parameter $\theta$
ranging over some set $\Theta$ in $\mathbb{R}.$ We assume that
there exists an estimator $\theta_n$ of the unknown parameter
$\theta,$ satisfying under $P_\theta,$  that the the random
variable $\sqrt{n}(\theta_n - \theta)$ converges in distribution
to $\mathcal{N}(0,\sigma^2)$ as $n\rightarrow \infty,$. It is well
known that $\frac {\sqrt{n}(\theta_n - \theta)}{\sigma_n} \sim T$,
where $T$ is a student distribution with $n -1$ degree of freedom
and $\sigma_n$ an estimator of $\sigma$. Its follows that with
confidence $(1 -\alpha) \times100\%$, the parameter $\theta$ is in
the interval $ IC_n,$ where, \\$
 IC_n=[\theta_n - \frac{\sigma_n~t_{\frac{\alpha}{2}}}{\sqrt{n}}~,~\theta_n +
\frac{\sigma_n~t_{\frac{\alpha}{2}}}{\sqrt{n}}] ~~\mbox{and}~~
t_{\frac{\alpha}{2}}  \mbox{~~satisfies to ~~} P(|T|\leq
t_{\frac{\alpha}{2}})=1 -\alpha.$\
 \noindent Throughout $S$ is a positive real, for any integer $n,$ let be
$N=[1 +n^{S + 1} ],$ where $"[~~]"$ is the integer part, we have
the following.
\begin{Proposition}\label{evaluation}
With confidence $(1 -\alpha) \times100\%$,
\begin{enumerate}
    \item  The random variable
$\theta_N - \theta$ converges in probability to $0$ with speed
$n^{{\beta} + \frac{1}{2}}$ where $\beta>0$.
    \item There exists a random
variable $R_n$ such that  $\sqrt{n}(\theta_n - \theta)=
\sqrt{n}(\theta_n - \theta_N) + R_n ,$ and  $R_n $ converges in
probability to 0 with speed $n^{\beta},$ with $\beta>0$.
\end{enumerate}
\end{Proposition}
In practice the Proposition \ref{evaluation} enables us to replace
the unknown quantity $\sqrt{n}(\theta_n - \theta)$ by the known
quantity $\sqrt{n}(\theta_n - \theta_N).$ The great advantage
about this replacing is  the convergence of the random variable
$R_n$ to 0 with speed $n^{\beta},$ with $\beta>0$ . Many
consequences are deduced from this previous evaluation. We expand
further some of them in the next section.
\section{Evaluation of the error and optimal tests in $AR(1)$ time series
model}\label{testuniv} \indent  Most of the cases,  the log
likelihood ratio was studied for a several classes of nonlinear
time series model which depend on unspecified parameter. The
replacing of this parameter by its estimator induces
asymptotically an no degenerate term in the expression of the
estimated central sequence. This latter alters the asymptotic
power. With the use of the evaluation \ref{evaluation}, we shall
preserving asymptotically the optimality. We focus our study on
the $AR(1)$ time series model with an extension to $ARCH$ models.
\noindent Let us recall some notations and assumptions used in the
remainder of this paper. We denote by $\Lambda_n$ the
log-likelihood ratio and we assume that $
\Lambda_n=\log\left(\frac{f_n}{f_{n,0}}\right),$ where
$f_{n,0}(\cdot)$ and $f_{n}(\cdot)$  denote the probability
densities of the random vector $(Y_1,\ldots,Y_n)$ corresponding to
the null hypothesis and the alternative hypothesis, respectively.
We also suppose that $\Lambda_n$ will be expressed as follows $
\Lambda_n=\sum_{i=1}^{n}\log(g_{n,i}).$ In a sequel, it will be
supposed that the error process \{$\epsilon_i$\} is independent
and identically distributed with zero mean, unit variance and a
positive density function $f$. For all $ x \in \mathbb{R}$, let be
$ M_f(x) =\frac{\dot{f}(x)}{f(x)},$\label{logderivée} where
$\dot{f}$ is the derivative of the function $f.$ Throughout
$\|\cdot\|_s$ is  the euclidian norm in $\mathbb{R}^s$.
\subsection{Testing in nonlinear time series contiguous to AR(1)
processes}\label{TEST1} \indent Consider the $s$-th order
 time series,\begin{eqnarray}
    Y_i =\theta Y_{i -1} + \alpha \,G(\textbf{Y}_i) + \epsilon_i\mbox{,} \quad |\theta| <1,\label{FIRSTAR1MODEL}
\end{eqnarray}where  $\textbf{Y}_i = \Big(Y_{i-1}, Y_{i-2},
\ldots,Y_{i-s}\Big)_{i\geq s},$ $\alpha$ a real parameter and $G$
a function with values in $\mathbb{R}$. \noindent In the sequel,
it will be assumed that the model  \ref{FIRSTAR1MODEL} is a
stationary and ergodic  time series with finite second moment. We
consider the problem of testing the null hypothesis $H_0:\alpha=0$
against the alternative hypothesis $H^{(n)}_1:\alpha
=n^{-\frac{1}{2}}\mbox{~~where~~} n\geq1.$ Clearly this testing
problem corresponds to test the linearity ($\alpha=0$) against a
no linearity ($\alpha =n^{-\frac{1}{2}}$) of the model
\ref{FIRSTAR1MODEL}. According to the previous notations and for
the constructing of a statistic test, we suppose that the
following conditions are satisfied.
\begin{description}
    \item $(L.1)$: $ \max_{1\leq i\leq n}| g_{n,i} - 1|=o_{P}(1).$
    \item $(L.2)$: There exists a positive constant ${\tau}^2$ such that   $ \sum_{i=1}^{n}(g_{n,i} - 1)^2 = {\tau}^2 + o_{P}(1).$
   \item $(L.3)$:  There exists a-$\mathcal{F}_{n}$  measurable
   $\mathcal{V}_n$ satisfying $ \sum_{i=1}^{n}(g_{n,i} - 1) = \mathcal{V}_n +
   o_{P}(1),$ where $\mathcal{V}_n\stackrel{\mathcal{D}}{\longrightarrow}
   \mathcal{N}(0,{\tau}^2).$
 \item $(A.1)$: There exists positive constants $\eta$ and $c$ such that for all $u$ with
    $\|u\|_s>\eta$, $G(u)\leq c\|u\|_s$.
   \item $(A.2)$: For a location family $\{f(\epsilon_i -c),~ -\infty
    <c<+\infty\}$, there exist a positive square integrable functions
    $\chi_1,$ $\chi_2$ and a positive constant $\delta$ such that, for all $\epsilon_i$ and
    $|c|<\delta,$ we have$
      \Big|\frac{{d}^k{f(\epsilon_i -c)}}{f(\epsilon_i)\,dc^k\,}\Big|
      \leq\chi_k(\epsilon_i)\mbox{,}\quad \mbox{for}\quad k=1,2.$
\end{description}
\indent It was established  in \cite{hb}[Theorem 1]  that the
conditions $(L.1),$ $(L.2)$ and $(L.3)$ imply under $H_0$ the
local asymptotic normality Lan of log-likelihood ratio
$\Lambda_n$, where,
\begin{eqnarray}
 \Lambda_n =\mathcal{V}_n(\theta) -\frac{{\tau}^2(\theta)}{2} +
 o_{P}(1)\label{lan}.
 \end{eqnarray}
Note that under $H_0$, the  residual $\epsilon_i$ and the
estimated residual $\hat{\epsilon}_i$ are given by,
\begin{eqnarray}
\epsilon_i= Y_i  - \theta Y_{i -1}, \mbox{~~and~~}
\hat{\epsilon}_i= Y_i - \theta_n Y_{i -1}. \label{residual}
 \end{eqnarray}
With the considering of an additional regularity conditions (A.1)
and (A.2), it was proved in \cite{hb}[Theorem 2] that the local
asymptotic normality of the time series model \ref{FIRSTAR1MODEL}
is established, the proposed test $T_{n}$ is the Neyman-Pearson
statistic which is given by the following equality\begin{eqnarray}
  T_{n} &=& I{\left\{{\frac{\mathcal{V}_{n}(\theta)}{\tau(\theta)}\geq Z(\alpha)}\right\}},
  \mbox{~where~} Z(\alpha) \mbox{~is the ~}(1 -\alpha)\mbox{-quintile of a standard
normal distribution~} \Phi(\cdot).\nonumber\\
\label{test}
\end{eqnarray}In this case, the central sequence is given by.\begin{eqnarray}
\mathcal{V}_{n}(\theta)=-\frac{1}{\sqrt{n}}\sum_{i=1}^{n}M_{f}(\epsilon_i)G(\textbf{Y}_i),\label{specifiedsequence1}\\
 \mbox{~~where~~} \tau^2=\tau^2(\theta)=\mathbf{E}({M^2_{f}}(\epsilon_{0}))\mathbf{E}(G^2(\textbf{Y}_0))\label{taux},\mbox{~~and  under~~}
 H_0,~~\mathcal{V}_{n}(\theta)\stackrel{\mathcal{D}}{\longrightarrow}\mathcal{N}(0,{\tau}^2).\noindent\end{eqnarray}
The asymptotic power of the test $T_{n}$ is derived and equal to
$1 - \Phi(Z(\alpha)-{\tau}^2 )$. Recall that when $\theta$ is
known, $T_{n}$ is asymptotically optimal. In practice the
parameter $\theta$ is unspecified, its estimation induces an
existence of an asymptotically non degenerate term . Therefore the
central and estimated central sequences are not equivalent. Based
on  the Proposition \ref{evaluation}, we shall obtain this
equivalence which implies the optimality of the constructed test.
Thus is expanded further in the next Subsection.\\
In the sequel, to aim to establish our results. We need  the
following
assumption.\\
 $(E.1):$ The functions
$x\longmapsto\dot{M}_{f}(x),$ $x \longmapsto \ddot{M}_{f}(x)$ and
 $x \longmapsto x\ddot{M}_{f}(x)$ are bounded, where ${M}_{f}$ is defined in
Section \ref{logderivée}, and $\dot{M}_{f}$, $\ddot{M}_{f}$ are the first and the second derivative of ${M}_{f}$ respectively.\\
Notice that, a large class of density functions satisfied $(E.1),$
we cite the gaussian distribution and the student distribution
with a degree of freedom $\geq 3.$
\subsubsection{Equivalence between the central sequences}\label{subequ}
\indent It will be assumed that the parameter $\theta$ is unknown,
let $\theta_n $ its consistent estimator. We suppose that the
 function  $\cdot \longrightarrow \mathcal{V}_n(\cdot)$ is
twice derivable on $\mathring{\Theta}$, we denote by
$\mathcal{\dot{V}}_n$ and $\mathcal{\ddot{V}}_n$ the first and
second derivative of $\mathcal{V}_n$ respectively. In order to
prove the equivalence between the central sequences, we need the
following assumptions.
\begin{description}
\item $(C.0)$: $\frac{1}{\sqrt{n}}
\mathcal{\dot{V}}_n(\theta_n)\neq 0.$
 \item $(C.1)$: $ \frac{1}{\sqrt{n}}
\mathcal{\dot{V}}_n(\theta_n)\stackrel{P}{\longrightarrow}c_{1}
\quad\mbox{as}\quad
 n\rightarrow\infty,$ where $c_{1}$ is some constant, such that $c_{1}\neq0.$
  \item $(C.2)$:  $\mathbb{E}(G(\textbf{Y}_0)^2)<\infty$.
   \item $(C.3)$: The fourth order moment of the stationary distributions of (\ref{FIRSTAR1MODEL}) are finite.
  \end{description}
According to the notations of the previous sections, we have the
following statement.
\begin{Proposition}\label{EQ1}

   With confidence $(1 -\alpha)
\times100\%$,
\begin{enumerate}
\item Under the
 hypothesis $H_0,$ $(A.1)$, $(A.2)$, ($(C.0)$-$(C.3)$) and
$(E.1)$ imply  that with $o_{P}(1)$, the sequences
 $\mathcal{V}_n(\theta)$  and   $\mathcal{W^S}_n(\theta_n)=(\theta_{[1 +n^S]} -\theta_n ) \mathcal{\dot{V}}_n(\theta_n)+
 \mathcal{V}_n(\theta_n)$ are equivalent.
 \item There exists a consistent estimator $\bar{\theta}_n$ of the
parameter $\theta,$ such that
$\mathcal{W^S}_n(\theta_n)=\mathcal{V}_n(\bar{\theta}_n),$ and
$\bar{\theta}_n$ is equal to $\theta_{[1 +n^S]}.$
\end{enumerate}
\end{Proposition}
Note that, the terms $(\theta_{[1 +n^S]} -\theta_n )
\mathcal{\dot{V}}_n(\theta_n)$ enables us to correct
asymptotically the error between the central sequence and its
estimator. From the Propositions \ref{EQ1}, we deduce that,
\begin{equation}
\mathcal{V}_n(\theta) = \mathcal{V}_n(\theta_{[1 +n^S]}) +
o_{P}(1)\label{lastequivalence1}.
\end{equation}
\subsubsection{Optimality of the test}\label{des2} \indent Now
we are ready to prove the optimality of the test \ref{test}. From
the equality \ref{lastequivalence1} the central sequences
$\mathcal{V}_n(\theta)$ and $\mathcal{V}_n(\theta_{[1 +n^S]})$ are
equivalent. Recall that, under
 the assumptions $(A.1)$ and $(A.2)$, the local asymptotic normality Lan is
established  and the power test function  is derived, see for more
detail \cite[Theorem 2]{hb}. With the replacing of $\theta$ by
$\theta_{[1 +n^S]}$ in the expressions \ref{test} and \ref{taux}
corresponding to the statistic test and the constant $\tau$ , we
obtain  the test $\bar{T}_n$ and $\tau(\theta_{[1 +n^S]})$
respectively,  then we have the following
theorem.\begin{Theorem}\label{optimaltest1} $(A.1)$, $(A.2)$,
$(C.0)$ and $(C.1)$ imply with confidence $(1 -\alpha)
\times100\%$ that the asymptotic power of $\bar{T}_n$ under
$H^n_1$ is equal  to $1 - \Phi(Z(\alpha)- \tau^2 (\theta_{[1
+n^S]})).$ Furthermore,  $\bar{T}_n$ is asymptotically optimal.
\end{Theorem}
\subsection{An extension to $ARCH$ model}\label{TEST2} \indent
Consider the following time series model with conditional
heteroscedasticity,
\begin{eqnarray}
    Y_i =\theta Y_{i -1} + \alpha \,G(\textbf{Y}_i) + \sqrt{1  + \beta B(\textbf{Y}_i)} \,\epsilon_i, \quad
    i\in\mathbb{Z},\label{model with conditional heteroscedasticity1}
\end{eqnarray}
where $\beta$ is real parameter, $B$ a function with values in
$\mathbb{R}$ and $\alpha$ and $G$ have the same
meaning as in \ref{FIRSTAR1MODEL}.\\
 First, let us recall some assumptions and notations.
It is supposed  that the model \ref{model with conditional
heteroscedasticity1} is ergodic and stationary, and  the
conditions $(B.1)$, $(B.2)$ and $(B.3)$ are satisfied, with,
\begin{description}
    \item $(B.1)$: The fourth order moment of the stationary distributions of (\ref{model with conditional heteroscedasticity1})  exists.
    \item $(B.2)$:  There exists a positive constants $\eta$ and $c$ such that, for all
    $u$,
    $\|u\|_s>\eta$,~~~ $B(u)\leq c{\|u\|_s}^2.$
   \item $(B.3)$: For a location family $\{b^{-1}f(\frac{\epsilon_i -a}{b}),~ -\infty
    <a<+\infty, ~b>0\}$, there exists a positive square integrable
function $\varphi(\cdot)$, and a strictly
    positive real $\varsigma$, where $\varsigma>\max(|a|,|b-1|)$, such that,
  $\left|\frac{{\partial}^2{b^{-1}f\left(\frac{\epsilon_i -a}{b}\right)}}{f(\epsilon_i)\,\partial
                            a^j\,
                            \partial b^k}\right|\leq\varphi(\epsilon_i),$
where $j$ and $k$ are two positive integers such that
                $j + k = 2.$
   \item $(B.4)$:  $\mathbb{E}(B(\textbf{Y}_0)^2)<\infty$.
 \end{description}
 We consider  the  problem of testing the
null hypothesis $H_0$ against the alternative hypothesis
$H^{(n)}_{1}$ such that $H_0 : \alpha=\beta=0,\mbox{~~and~~}
H^{(n)}_{1} :\alpha=\beta={n}^{-\frac{1}{2}}.$ Note that when $n$
is large, we have $\sqrt{1 + n^{-\frac{1}{2}} B(\textbf{Y}_i
)}\simeq  1 + \frac{n^{-\frac{1}{2}}}{2} B(\textbf{Y}_i)=1
  +n^{-\frac{1}{2}} L(\textbf{Y}_i).$ Under the conditions $(A.1)$, $(B.1)$, $(B.2)$, and $(B.3)$, the
local asymptotic normality was established in \cite[Theorem
4]{hb}, an efficient test is obtained and its power function is
derived. In this case, we have  the following equalities.
\begin{eqnarray}
\mathcal{V}_{n}(\theta)=-\frac{1}{\sqrt{n}}\left\{\sum_{i=1}^{n}M_{f}(\epsilon_i)G(\textbf{Y}_i)
+ \sum_{i=1}^{n}(1 + \epsilon_i
M_{f}(\epsilon_i))L(\textbf{Y}_i)\right\},\label{secondcentral}\\
 {\tau}^2 = I_0 \mathbf{E}\left(G(Y_0)
        \right)^2
+ \frac{(I_2 - 1)}{4} \mathbf{E}\left(L(Y_0) \right)^2 + I_1
\mathbf{E}\left(G(Y_0))L(Y_0)\right),\nonumber\\
\mbox{~~where~~}
I_j=\mathbf{E}\Big({\epsilon}^j_{0}{M^2_{f}}(\epsilon_{0})\Big),
\mbox{~and~} j=0,1,2.\label{secondtaux}
 \end{eqnarray}
The Neyman-Pearson statistic is used again with the replacing of
the central sequence \ref{specifiedsequence1} by
\ref{secondcentral} and the constant \ref{taux} by
\ref{secondtaux}. In order to prove the optimality, we first study
 the equivalence between the central and the
estimated central sequences, we
obtain,\begin{Proposition}\label{EQ3}
 With confidence $(1 -\alpha)
\times100\%$,\begin{enumerate} \item $(A.1)$, $((B.1)-(B.4))$,
$(C.0)$ , $(C.1)$ and $(E.1)$ imply under the hypothesis $H_0,$
that with $o_{P}(1)$, the sequences
 $\mathcal{V}_n(\theta)$  and   $\mathcal{W^S}_n(\theta_n)=(\theta_{[1 +n^S]} -\theta_n ) \mathcal{\dot{V}}_n(\theta_n)+
 \mathcal{V}_n(\theta_n)$ are equivalent.
\item There exists a consistent estimator $\bar{\theta}_n$ of the
parameter $\theta,$ such that
$\mathcal{W^S}_n(\theta_n)=\mathcal{V}_n(\bar{\theta}_n),$ and
$\bar{\theta}_n$ is equal to $\theta_{[1 +n^S]}.$
\end{enumerate}
\end{Proposition}
It follows from the Propositions \ref{EQ3} that, $
\mathcal{V}_n(\theta) = \mathcal{V}_n(\theta_{[1 +n^S]}) +
o_{P}(1).$
\subsubsection{Optimality of the test}\label{des3} \noindent From
the last previous equality, it follows that the central sequences
$\mathcal{V}_n(\theta)$ and $\mathcal{V}_n(\theta_{[1 +n^S]})$ are
equivalent. With the replacing of $\theta$ by $\theta_{[1 +n^S]}$
in the expressions \ref{test} and \ref{taux} corresponding to the
statistic test and the constant $\tau$, we obtain  the test
$\bar{T}_n$ and $\tau(\theta_{[1 +n^S]})$ respectively,  then we
have the following theorem.\begin{Theorem}\label{optimaltest2}
$(A.1)$, $(B.1)$, $(B.2)$, $(B.3)$, $(C.0)$, $(C.1)$ and $(E.1)$
imply with confidence $(1 -\alpha) \times100\%$ that the
asymptotic power of $\bar{T}_n$ under $H^n_1$ is equal  to $1 -
\Phi(Z(\alpha)- \tau^2 (\theta_{[1 +n^S]})).$ Furthermore,
$\bar{T}_n$ is asymptotically optimal.
\end{Theorem} The generalization of these results is effective under  some
assumptions. The study of this possibility  is detailed in the
next section.
\section{Generalization of the results }\label{general}
\indent Moving from the testing in nonlinear time series
contiguous to $AR(1)$ processes to testing in $ARCH$ models was
both easy and obvious. But, the more important assumption was
naturally the establish of local asymptotic normality Lan of the
log-likelihood ratio which plays an important role. Under this
fundamental assumption, we shall present in this Section an
extension of our results  in a class of nonlinear time series
model with high order $d$ described in the following. Let $Y_{i}$
be a sequence  of stationary and ergodic random  with finite
second moment such that for all $i \in \mathbb{Z}$. We consider
the class of stochastic models
\begin{eqnarray}
    Y_i =\psi_1(\textbf{Y}_{i}) +\psi_2(\textbf{Y}_{i})\,\epsilon_i, \quad
    i\in\mathbb{Z},\label{modelprincipal}
\end{eqnarray}where, for given non negative  integer $d$, the random
vectors $  \textbf{Y}_{i-1} $  is equal to $
\textbf{Y}_{i}=(Y_{i-1}, Y_{i-2}, \ldots, Y_{i-d})_{i\geq d}$, the
$\epsilon_{i}$'s are centered i.i.d. random variables with unit
variance and density function $f(\cdot)$, such that for each $i
\in \mathbb{Z}$, $\epsilon_{i}$ is independent of the filtration
$\mathcal{F}_{i} = \sigma(Y_j, j \leq i),$. We consider the
problem of
 testing whether the bivariate vector of
functions $(\psi_1(\cdot),\psi_2(\cdot))$ belongs to a given class
of parametric functions or not. More precisely, let\\
$\mathcal{PF}=\left\{\left(m({\theta},\cdot),
\sigma({\rho},\cdot)\right) ,~
({\theta}^\prime,{\rho}^\prime)^\prime \in \Theta_1 \times
\Theta_2 \right\},$~~ $ \Theta_1 \times
\Theta_2\subset\mathbb{R}^s\times\mathbb{R}^t,$
 $\mathring{\Theta}_1\neq \emptyset,$    $\mathring{\Theta}_2\neq \emptyset,$
where for all set $A$, $\mathring{A}$ denotes the interior of the
set $A$ and the script ``$~~^\prime~~ $'' denotes the transpose,
$s$ and $t$ are two positive integers, and each one of the two
functions $m({\theta},\cdot)$ and $\sigma({\rho},\cdot)$ has a
known form such that $\sigma({\rho},\cdot)>0.$ \noindent For a
sample of size $n$, we derive a test of $H_0:
\left[(\psi_1(\cdot),\psi_2(\cdot))\in \mathcal{PF}\right]
\mbox{~~~against~~~}
H_1:\left[(\psi_1(\cdot),\psi_2(\cdot))\notin\mathcal{PF}\right].$
Observe that the null hypothesis $H_0$ is equivalent to $
  H_0:[(\psi_1(\cdot),\psi_2(\cdot)] =
  \Big(m(\theta,\cdot),\sigma(\rho,\cdot)\Big),\label{principal null
  hypothesis}$
and  the alternative hypothesis $H_1$ is equivalent to $
H_1:[(\psi_1(\cdot),\psi_2(\cdot)] \neq
\Big(m(\theta,\cdot),\sigma(\rho,\cdot)\Big),\mbox{~~for some
~~}(\theta^\prime,\rho^\prime)^\prime \in \Theta_1 \times
\Theta_2. $ For a specified  real functions  $G(\cdot)$ and
$L(\cdot)$, we focus our study on  the following form of the the
alternative hypothesis $H^{(n)}_1,$ such that,$
H^{(n)}_1:[(\psi_1(\cdot),\psi_2(\cdot)]= \Big(m(\theta,\cdot)
+\,n^{-\frac{1}{2}}G(\cdot),\sigma(\rho,\cdot)+\,
n^{-\frac{1}{2}}L(\cdot)\Big),\label{principal alternative
hypothesis} \mbox{~~where~~} n\geq1.$ Note that under $H_0$,
\ref{modelprincipal} is equal to\begin{eqnarray}
    Y_i =m(\theta,\textbf{Y}_{i}) + \sigma(\rho,\textbf{Y}_{i})\,\epsilon_i, \quad
    i\in\mathbb{Z},\label{modelprincipal2}
\end{eqnarray} and under $H^{(n)}_1$, \ref{modelprincipal} is equal to
\begin{eqnarray}
 Y_i =m(\theta,\textbf{Y}_{i})+\,n^{-\frac{1}{2}}G(\textbf{Y}_{i}) +
 [\sigma(\rho,\textbf{Y}_{i})+\,n^{-\frac{1}{2}}L(\textbf{Y}_{i})]
 \,\epsilon_i. \quad
    i\in\mathbb{Z}\label{modelprincipal3}
\end{eqnarray}
For the stating of the local asymptotic normality of the model
\ref{modelprincipal2}, we refer to the paper of \cite{fl} which is
an extension of  the work of \cite{hb} . Let us recall the
notations and assumptions used in. Let be  $\varphi_x$  a function
such that for each $(a,b),~ {\varphi}_x(a,b)={b}^{-1}f
\big(b^{-1}(x-a)\big)$, where $b \neq0$. The following regularity
conditions were required.
\begin{description}
\item (D.1): There exists a positive measurable function $M$
satisfying $\mathbb{E}|M|^{1+\gamma}<\infty$ for some $\gamma>0,$
and some $\delta>0$ such that for $|a|<\delta$ and $|b -1|<\delta$
we have ~~ $\Big|\frac{1}{f(x)} \frac{\partial^2}{\partial a^j
\partial b^k} \varphi_x(a,b)\Big | \leq
    M(x)$  for a positive integers $j$ and $k$ with j+k=2.
\item (D.2): There exists $\gamma'>0$ such that
$\mathbb{E}\Big|\frac{L(\textbf{Y}_d)}{\sigma(\rho,\textbf{Y}_{d-1})}\Big|^{2
+ \gamma'}
<\infty,$~~and~~$\mathbb{E}\Big|\frac{G(\textbf{Y}_d)}{\sigma(\rho,\textbf{Y}_{d-1})}\Big|^{2
+ \gamma'} <\infty.$
 \item (D.3): $\mathbb{E}\Big|
M_{f}(\epsilon_d){\epsilon_d}^k  \Big|^{2 + \gamma''} <\infty $
for some $\gamma''>0$ and k =0 , 1.
 \item (D.4):
  \begin{description}
        \item (D.4.1):~$\mathbb{E}\left\{ M_{f}(\epsilon_{0})\right\}=0.$
     ~~(D.4.2):~$\mathbb{E}\Big\{\epsilon_{0}M_{f}({\epsilon_{0}})
     \Big\}=-1.$~~(D.4.3):~$\mathbb{E}\left \{ \dot{M}_{f}(\epsilon_{0}) +
{M^2_{f}}(\epsilon_{0})\right\}=0.$ \item~~ (D.4.4):~
$\mathbb{E}\left \{ \epsilon_{0}(\dot{M}_{f}(\epsilon_{0}) +
{{M^2_{f}}}(\epsilon_{0}))\right\}=0.$~~  (D.4-5):~$
\mathbb{E}\left \{ \epsilon^2_{0}(\dot{M}_{f}(\epsilon_{0}) +
{{M^2_{f}}}(\epsilon_{0}))\right\}=2.$
 \end{description}
\end{description}
From \cite{fl}[Theorem 2.1], when the conditions ((D.1)-(D.4)) are
fulfilled, under $H_0$, the local asymptotic normality of the
model \ref{modelprincipal2} is established  with a same expression
as \ref{lan}, where

\begin{eqnarray}
\mathcal{V}_{n}(\theta)&=&-\frac{1}{\sqrt{n}}\left\{\sum_{i=1}^{n}M_{f}(\epsilon_i)G(\textbf{Y}_i)
+ \sum_{i=1}^{n}(1 + \epsilon_i M_{f}(\epsilon_i))L(\textbf{Y}_i)\right\},\label{secondcentral1}\\
  {\tau}^2 &=&  I_0 \mathbb{E}\left(\frac{G(\textbf{Y}_d)}
        {{\sigma}(\rho,\textbf{Y}_d)}\right)^2
+(I_2 - 1) \mathbb{E}\left(\frac{L(\textbf{Y}_d)} {{\sigma}(\rho,
       \textbf{Y}_d)}\right)^2 + 2I_1 \mathbb{E}\left(\frac{G(\textbf{Y}_d)L(\textbf{Y}_d)} {{\sigma^2}(\rho,
       \textbf{Y}_d)}\right),\nonumber\\ \label{tauxgeneral}
\mbox{~and~} I_j=\mathbb{E}\Big({\epsilon}^j_d{M^2_{f}}(\epsilon_d)\Big),~~~~~~~j\in\{0,1,2\}.\nonumber
\end{eqnarray}
Many specifications of $m(\theta,\cdot)$ and $\sigma(\rho,\cdot)$
show that \ref{modelprincipal2}  embodies a large variety of
famous time series models. Namely, we cite the $AR(m)$ time series
model, where $m\geq1$. The obtaining of the optimality remains the
more important objective. Based on this last version of Lan, it is
possible to obtain an extension in $AR(m)$ time series model. The
details are presented in the next Subsection.
\subsection{Testing in $AR(m)$ time series model}\label{TEST3}
Consider the following $AR(m)$ time series model.\begin{eqnarray}
   Y_i =\sum_{j=1}^{m}\theta_j  Y_{i-j}~~  + \epsilon_i, \quad
\mbox{where} \quad \sum_{j=1}^{m} |\theta_j|  <1.\label{ARm model}
\end{eqnarray}
It will assumed that the model  \ref{ARm model} is stationary and
ergodic with finite second and fourth moments,  in this case, and
by the comparison with \ref{modelprincipal2}, we have. $
m(\theta,\textbf{Y}_i)= \sum_{j=1}^{m}\theta_j  Y_{i-j},
\mbox{~~and~~} \sigma(\rho,\textbf{Y}_i)=1.$  We denote by
$\theta_{n}^\prime=\Big({\theta}_{n,1},\dots,\theta_{n,m})$ the
$\sqrt{n}$-consistence estimator of the unknown parameter
$\theta^\prime=\Big({\theta}_1,\dots,\theta_m)$. For all $x$, we
suppose that, the function $x\rightarrow \mathcal{V}_{n}(x)$ is
twice differentiable on $\mathring{\Theta}$.\\ We denote by $
\nabla \mathcal{V}_{n}(\cdot)^\prime= (\frac{\partial
\mathcal{V}_{n}(\cdot)}{\partial{\theta}_{1}},\dots,\frac{\partial
\mathcal{V}_{n}(\cdot)}{\partial
 {\theta}_{m}}), \mbox{~~and~~}
\partial^2\mathcal{V}_{n}(\cdot)(\theta,\cdot)=\Big(\frac{\partial^2\,\mathcal{V}_{n}(\cdot)(\theta,\cdot)}{\partial\theta_{i}\partial\theta_{j}}\Big)_{1\leq i ,j \leq
m }$, where $\partial^2\mathcal{V}_{n}(\cdot)$ is the hessian
matrix of $\mathcal{V}_{n}(\cdot)$ in $\theta$. At first, with the
use of the equality \ref{secondcentral}, we specify the link
between the estimated central and the central sequences. We have
the following statement.
\begin{Proposition}\label{link}
((D.1)-(D.4)) and (E.1) imply that
\begin{eqnarray}\mathcal{V}_n(\theta) - \mathcal{V}_n(\theta_n)=\nabla\mathcal{V}_n(\theta_n)\cdot(\theta  -\theta_n)
~+ o_p(1),\mbox{~~ where~~} "\cdot"  \mbox{~~ is a inner
product}.\label{linkgeneralcase}\end{eqnarray}
\end{Proposition}
In general case, the term
$\nabla\mathcal{V}_n(\theta_n)\cdot(\theta -\theta_n)$ is
asymptotically no degenerate. In order to avoid the effect of this
error, we need to give an evaluation similar as in the Section
\ref{evaluation univa}. Since the parameter is multivariate, we
need to work with simultaneous confidences intervals. Consider
again the model \ref{ARm model} and assume that  the following
assumption are satisfied. $$(R.1):\sqrt{n}(\theta_n -
\theta)\stackrel{\mathcal{D}}{\longrightarrow}
   \mathcal{N}_m(0,\Sigma),\mbox{~~~where~} \Sigma \mbox{~is a~} m\times m \mbox{~ matrix~}.\label{regularity}
$$
Note that the condition \ref{regularity} is fulfilled for the LSE
estimators. But, we require that the characteristic equation for
the model \ref{ARm model} has any solution in the unit disc.
\noindent In order to construct a simultaneous confidence
intervals for the parameters ${(\theta_ j)}_{j=1,\dots,m.}$, we
use Delta method, then we obtain the following statement.
\begin{Proposition}\label{delta method and confidence intervals}
Under $(R.1),$ with confidence $(1 -\alpha) \times100\%,$ we have,
for each $j\in \{1,\dots,m\},$ $$ \theta_j \in [\theta_{n,j} -
\frac{\tilde{\Sigma}_{j,j}~t_{\frac{\alpha}{2}}(n-m)}{\sqrt{n}}~;~\theta_{n,j}
+
\frac{\tilde{\Sigma}_{j,j}~t_{\frac{\alpha}{2}}(n-m)}{\sqrt{n}}],
\mbox{~~where~~} \tilde{\Sigma} \mbox{~~is the observed  matrix }
.$$
\end{Proposition}
We can see that this Proposition implies a generalization of the
Proposition \ref{evaluation}, thus is given by the following
statement.\begin{Lemma}\label{evaluationm } With confidence $(1
-\alpha) \times100\%$, and for each $j\in\{1,\dots, m\}$, the
random variable $\theta_{N,j} - \theta_j$ converges in probability
to $0$ with speed $n^{{\beta}_j + \frac{1}{2}}$ where $\beta_j>0$.
\end{Lemma}
\noindent With the use of an analogous methodology as in
Subsection \ref{subequ}, we shall state the following results.
\begin{Proposition}\label{EQ5} With confidence $(1 -\alpha)
\times100\%$,
\begin{enumerate}
  \item  ($(D.1)$-$(D.4)$), $(C.0)$, $(C.1)$ and
  $(E.1)$ imply under the
 hypothesis $H_0,$ that with $o_{P}(1)$, the sequences
 $\mathcal{V}_n(\theta)$  and   $\mathcal{W^S}_n(\theta_n)=\nabla(\theta_n)\cdot(\theta_{[1 +n^S]} -\theta_n ) +
 \mathcal{V}_n(\theta_n)$ are equivalent.
 \item  There exists a consistent estimator $\bar{\theta}_n$ of the
parameter $\theta,$ such that
$\mathcal{W^S}_n(\theta_n)=\mathcal{V}_n(\bar{\theta}_n).$
\end{enumerate}
\end{Proposition}
 The combination of the two Propositions \ref{link} with \ref{EQ5}
 implies that,~~ $\mathcal{V}_n(\theta) = \mathcal{V}_n(\bar{\theta}_n) ~+~
 o_p(1).$\\
With following the same  reasoning as in Subsections \ref{des2}
and \ref{des3}, we deduce the following statement.
\begin{Theorem}\label{optimaltest3}
 $(C.0)$, $(C.1)$, $((D.1)-(D.4))$ and $(E.1)$ imply with
confidence $(1 -\alpha) \times100\%$ that the asymptotic power of
$\bar{T}_n$ under $H^n_1$ is equal  to $1 - \Phi(Z(\alpha)- \tau^2
(\bar{\theta}_n)).$ Furthermore,  $\bar{T}_n$ is asymptotically
optimal.
\end{Theorem}
\section{Simulations}\label{simul}
\subsection{Testing in nonlinear time series contiguous to AR(1)
processes}\label{SIMULtest1}
 \noindent To aim  to study the performance of our
evaluation, simulations are carried out. In order to test the null
hypothesis $H_0:\alpha=0$ against the alternative hypothesis
$H^{(n)}_1:\alpha =n^{-\frac{1}{2}},$ corresponding to the ergodic
$AR(1)$ time series model with parameter $\theta$, where
 $
    Y_i =\theta Y_{i -1} + \alpha \,G(\textbf{Y}_i) + \epsilon_i\mbox{,} \quad |\theta| <1,
  ~~ \epsilon_i \mbox{~~are centred i.i.d
with unit variance and } \\ \epsilon_{0}
\stackrel{\mathcal{D}}{\longrightarrow} \mathcal{N}(0,1).$ The
true value of the parameter $\theta$ is fixed at $0.6$ and the
sample sizes are  fixed at $n = 30, 49, $ and $52$. The values of
$S$ are fixed at $ 1$. Based on the observations $Y_1 \dots Y_n,$
we estimate the parameter $\theta$ by the least square estimator
$\theta_n.$ We denote the estimator $\bar{\theta}_n =\theta_[1
+n^{S + 1}],$ by the S-estimator. Based on the inequality $a^2 +
b^2 \geq 2ab$, we shall choose $G:x \longrightarrow \frac{1}{1 +
x^2}$ for satisfying the condition $(A.1).$ All these estimation
were made upon $m=50$ replicates. Recall that  in this case, we
have $\mathbb{E}({\epsilon_{i}})=0,$ \quad
$\mathbb{E}({\epsilon^2_{i}})=1,$ \quad  and
$\mathbb{E}({\epsilon^4_{i}})=3.$\\
The residual $\epsilon_i= Y_i  - \theta Y_{i -1}$ is estimated by
$\tilde{\epsilon}_i= Y_i  - \theta_n Y_{i -1}$ and
$\bar{\epsilon}_i= Y_i  - \bar{\theta}_n Y_{i -1}$ corresponding
to the LSE and S-estimator respectively. With the replacing in the
expression  \ref{taux} of $\epsilon_i$ by the $\tilde{\epsilon}_i$
and $\bar{\epsilon}_i$ respectively, we obtain,
$\tilde{\tau}^2=\tau^2(\theta_n)=\mathbf{E}({M^2_{f}}(\tilde{\epsilon}_{0}))\mathbf{E}(G^2(\textbf{Y}_0))\mbox{~~and~~}
\bar{\tau}^2=\tau^2(\theta_N)=\mathbf{E}({M^2_{f}}(\bar{\epsilon}_{0}))\mathbf{E}(G^2(\textbf{Y}_0)).$
According to the notations  and assumptions of the Subsection
\ref{TEST1}, we represent simultaneously the power functions of
the test \ref{test} and the power functions after replacing the
parameter $\theta$ by  LSE and the S-estimator. We obtain the
upper representations. One remark, that the power function
corresponding to the S-estimator is closer to the true power
function.

\begin{figure}[h!]
\includegraphics[scale=0.4]{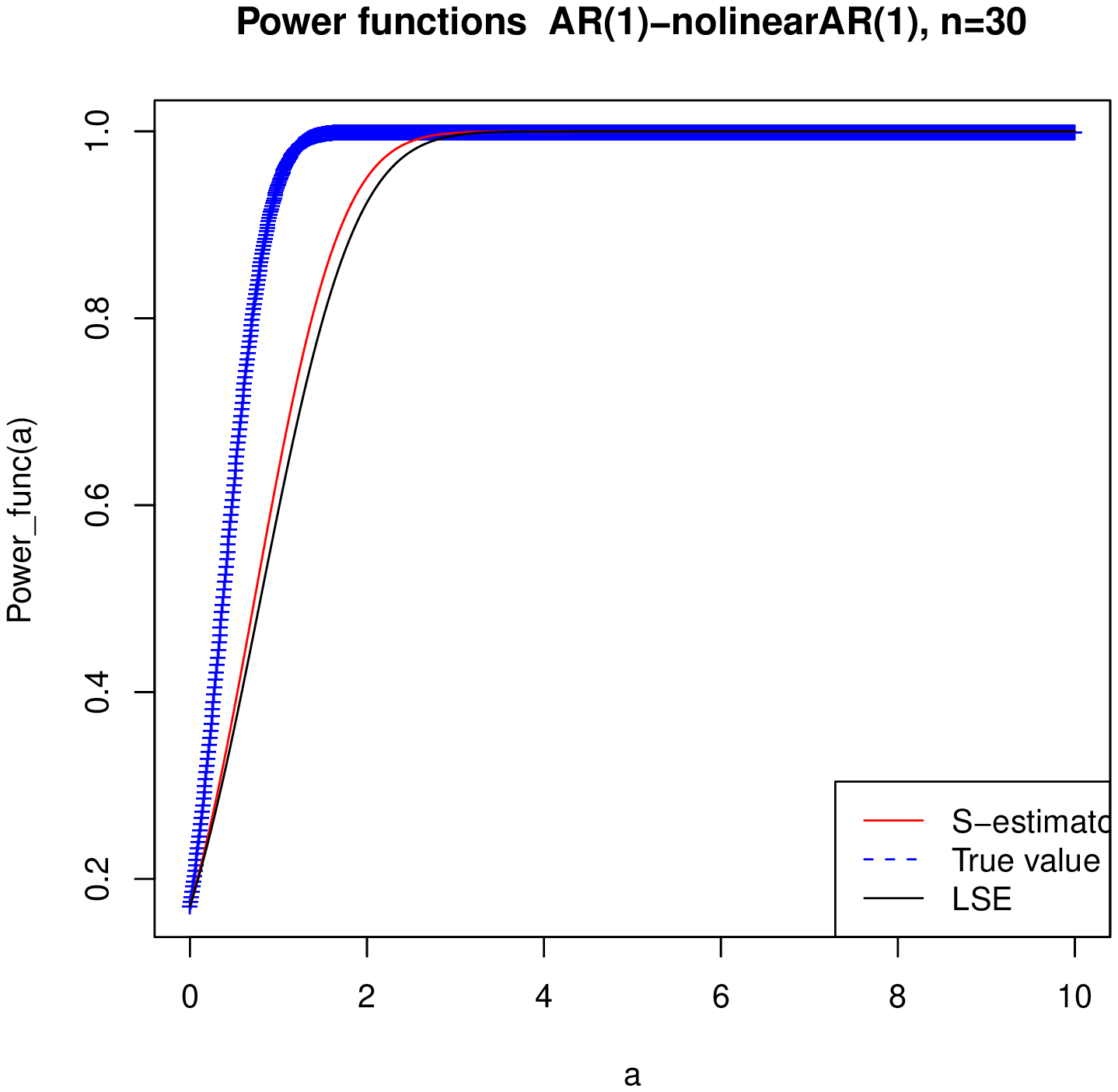}
\includegraphics[scale=0.3]{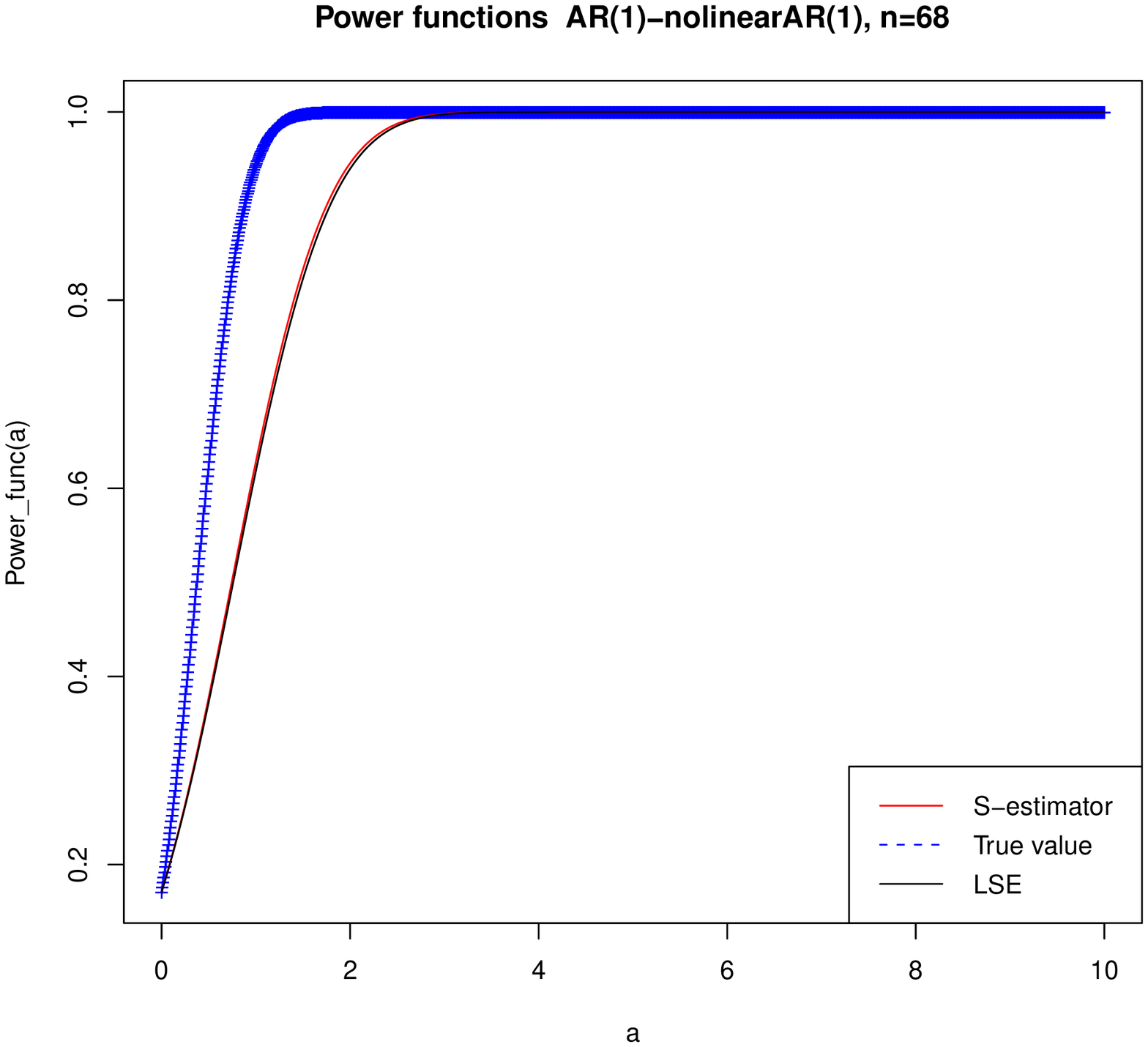}
\includegraphics[scale=0.3]{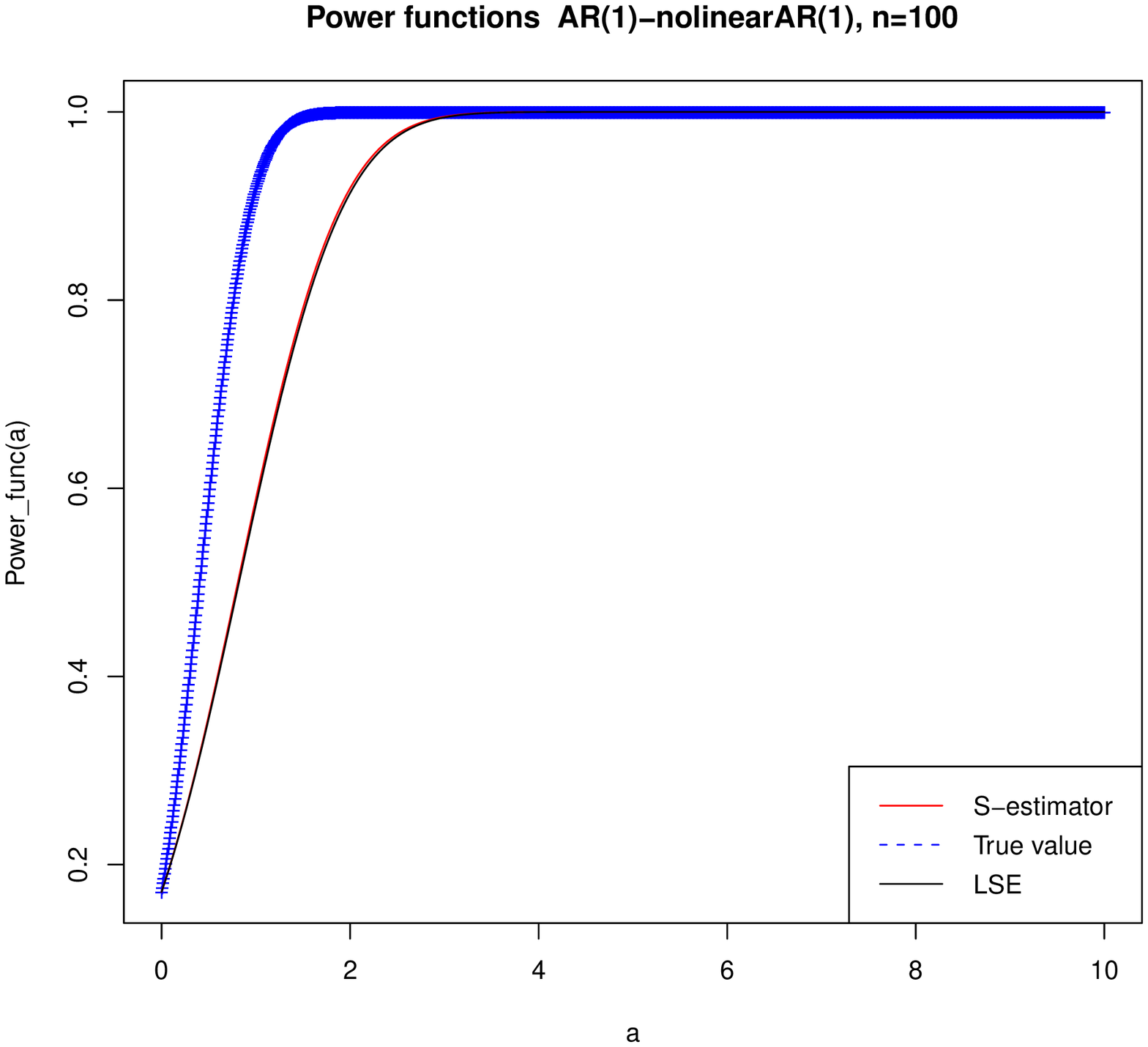}
\end{figure}

\subsection{Testing in $ARCH$ processes}\label{SIMULtest2}
 \indent  Consider again  the series model \ref{model with conditional heteroscedasticity1}.
 For the studying of  testing's problem between null hypothesis
$H_0$ and the alternative hypothesis $H^{(n)}_{1}$, where $H_0 :
\alpha=\beta=0,\mbox{~~and~~} H^{(n)}_{1}
:\alpha=\beta={n}^{-\frac{1}{2}},$ we use again the Neyman-Pearson
statistic test. According to the notations and assumptions of the
Subsection \ref{TEST2}, we made a same work as the last previous
Subsection. Note that in this case,the central sequence
$\mathcal{V}_{n}(\theta)$ and the constant $\tau$ are given by the
equalities \ref{secondcentral} and \ref{secondtaux} respectively.
Also, $L(x)=G(x):x \longrightarrow \frac{1}{1 + x^2}$ and the
sample sizes are  fixed at $n = 30, 68, $ and $100$. All those
simulations are made upon $m=100$ replicates. We obtain the
following figures with a same conclusion as in Subsection
\ref{SIMULtest1}.
\begin{figure}[h!]
\includegraphics[scale=0.3]{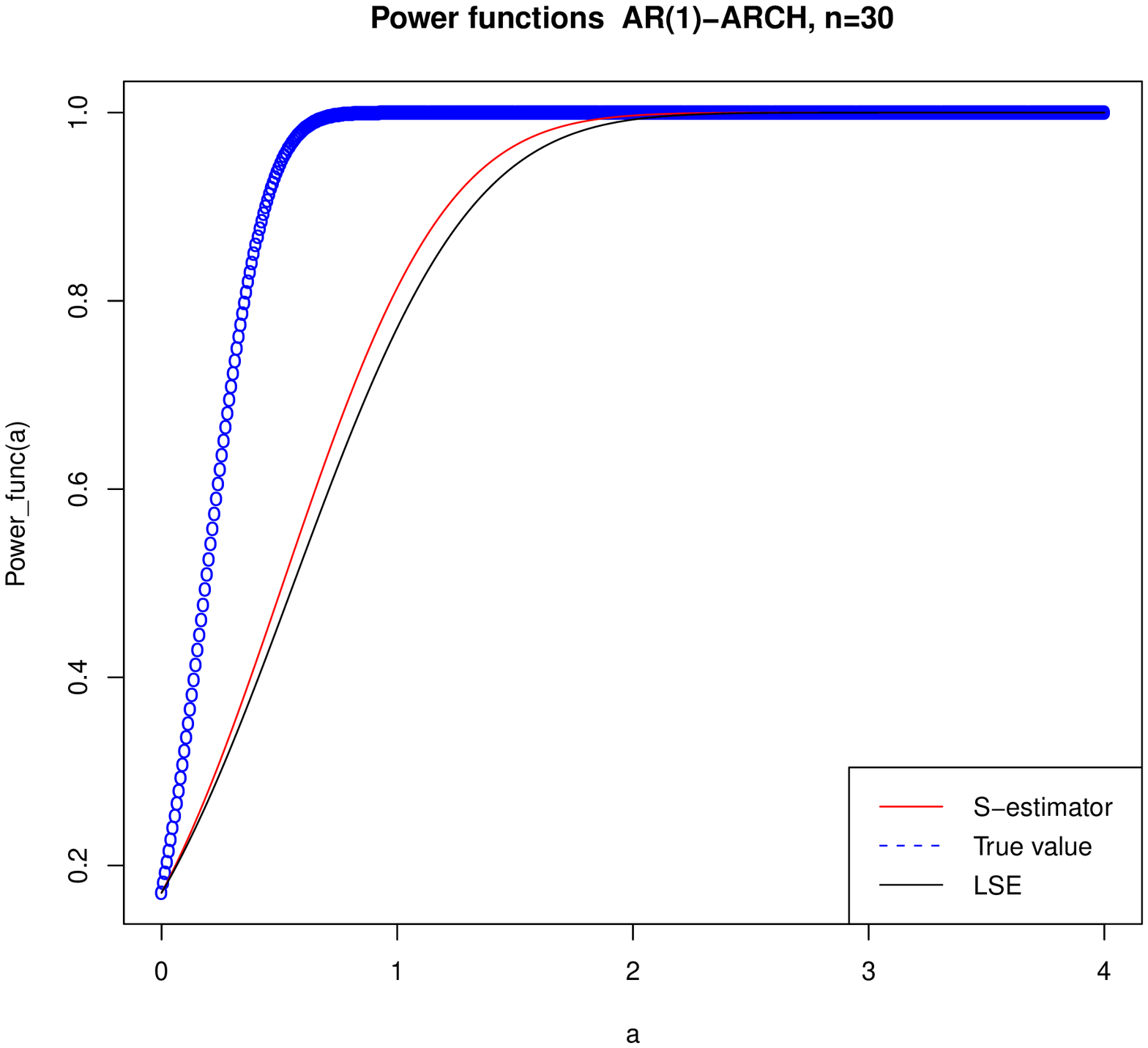}
\includegraphics[scale=0.3]{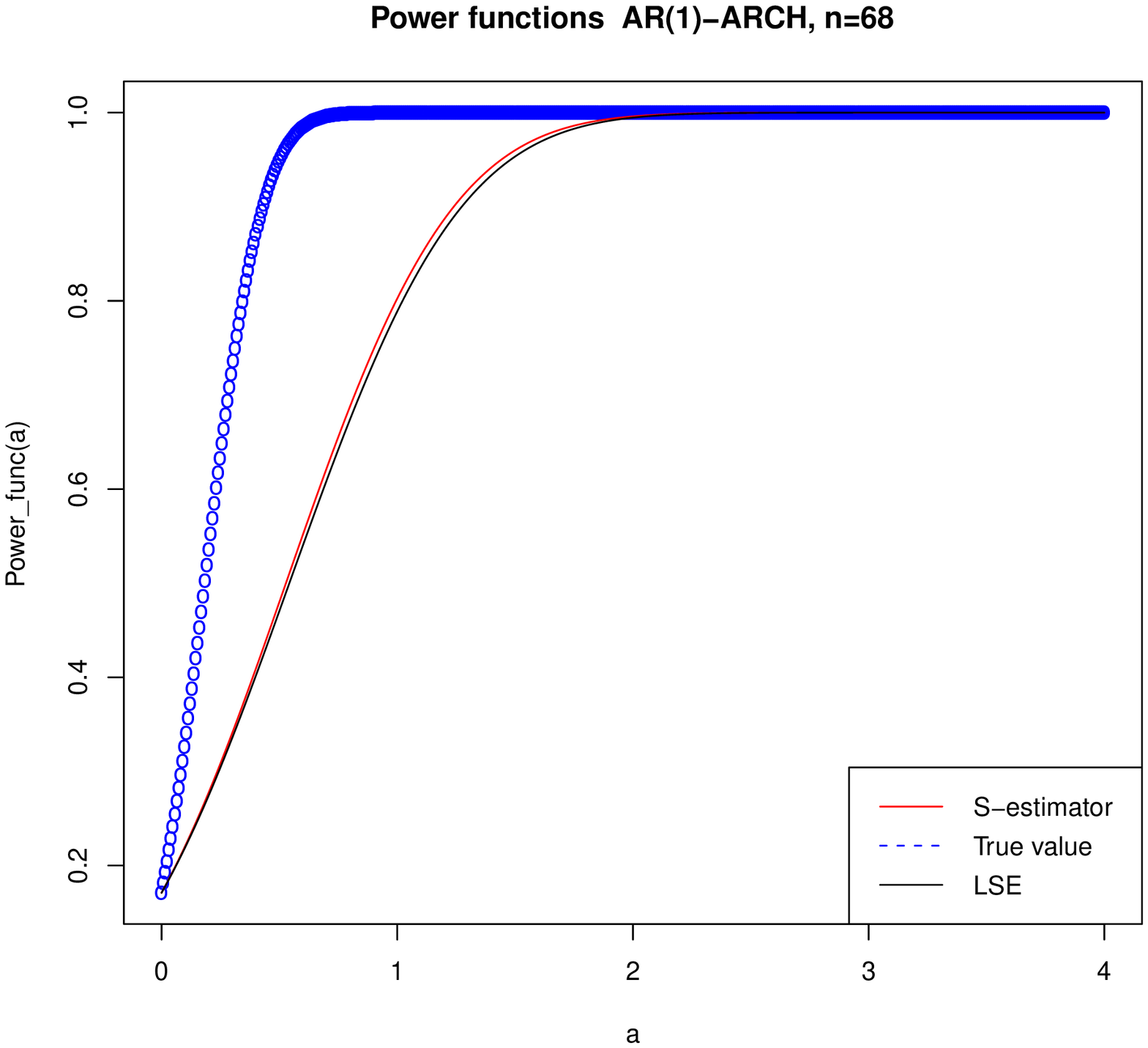}
\includegraphics[scale=0.3]{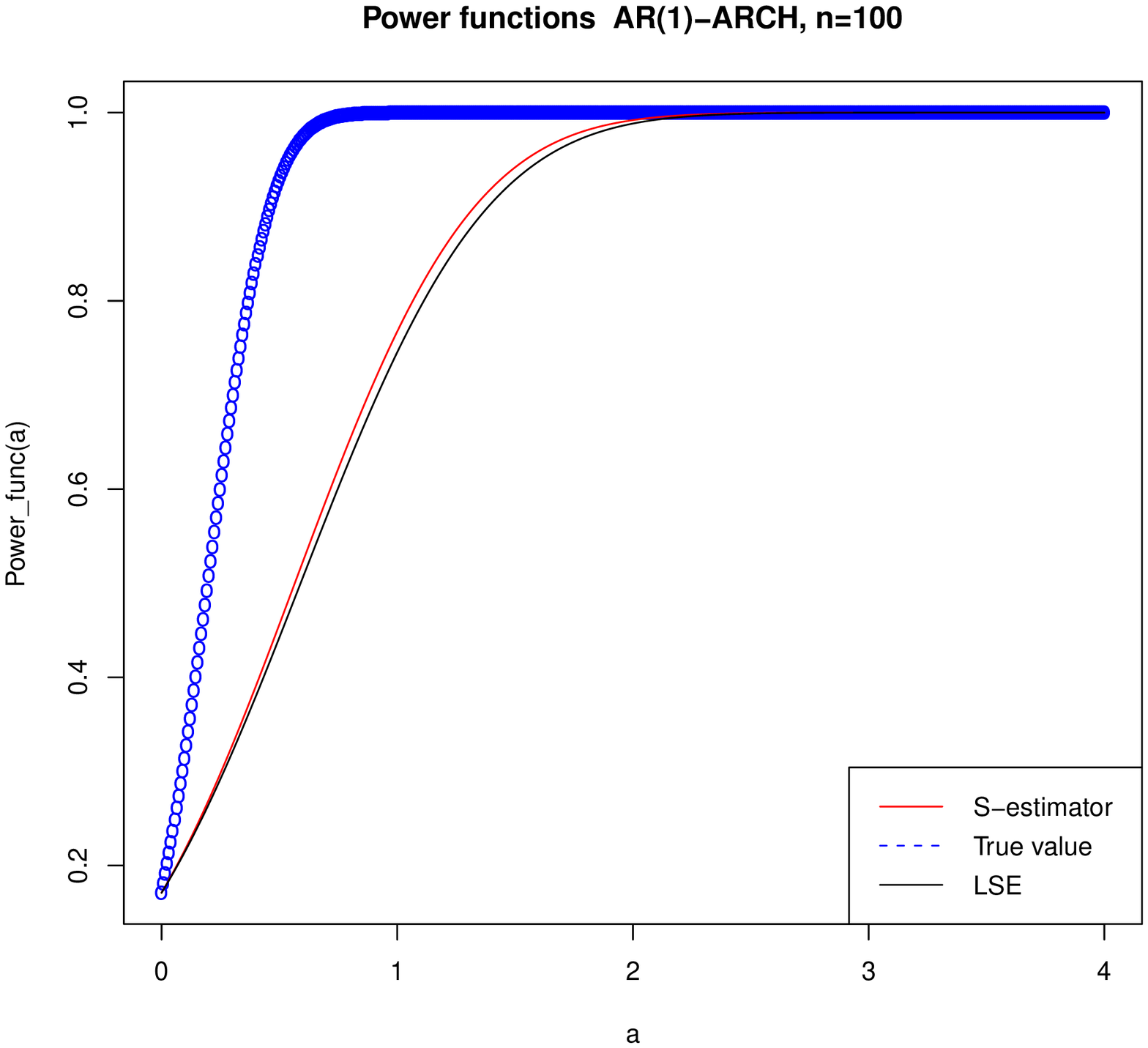}
\end{figure}
\section{Proof of the results}\label{demonstration}
\noindent {\bf Proof of the Proposition \ref{evaluation}}
\begin{enumerate}
    \item \label{1}\noindent With confidence $(1 -\alpha) \times100\%$, the unknown
parameter $\theta$ is in the confidence interval $IC_N,$
therefore,
$$|\theta_N - \theta|\leq |2t_{\frac{\alpha}{2}}\sigma_N
|\times \frac{1}{\sqrt{[1+n^{1+{S}}}]}.$$ Since,
$$ n^{1+{S}}\leq[1+n^{1+{S}}]\leq 2
+ n^{1+{S}},$$
it follows that,
$$ \frac{1}{\sqrt{2+n^{1+{S}}}}\leq \frac{1}{\sqrt{[1+n^{1+{S}}]}}\leq \frac{1}{\sqrt{n^{1+{S}}}}.$$
We obtain,
$$|\theta_N - \theta|\leq |2t_{\frac{\alpha}{2}}\sigma_N
|\times \frac{1}{\sqrt{n^{1+{S}}}}.$$
 Because the
quantity $|2t_{\frac{\alpha}{2}}\sigma_N |$ is bounded and with
the choice of $\beta=\frac{S}{4},$ we obtain,
$$n^{{\beta} +
\frac{1}{2}}\times|\theta_N - \theta|\leq
|2t_{\frac{\alpha}{2}}\sigma_N |\times
n^-{\frac{S}{4}}\stackrel{\mathbb{P}}{\longrightarrow}0.$$
    \item \noindent Consider the following decomposition, $$\sqrt{n}(\theta_n
- \theta)= \sqrt{n}(\theta_n - \theta_N) + \sqrt{n}(\theta_N  -
\theta)= \sqrt{n}(\theta_n - \theta_N) + R_n.$$ From \ref{1} ,
$R_n$ converges in probability to $0$ with speed $n^\beta.$
\end{enumerate}
\noindent {\bf Proof of the Proposition \ref{EQ1}}
 \begin{enumerate}
    \item By Taylor expansion of the function $\mathcal{V}_n$ with
order $2$ at $\theta_n,$
\begin{equation}
\mathcal{V}_n(\theta) - \mathcal{V}_n(\theta_n) = (\theta -
\theta_n)\mathcal{\dot{V}}_n(\theta_n)  +\frac{1}{2}(\theta -
\theta_n)^2 \mathcal{\ddot{V}}_n(\tilde{\theta}_n)=
\sqrt{n}(\theta - \theta_n)\times
\frac{\mathcal{\dot{V}}_n(\theta_n)}{\sqrt{n}} +\frac{1}{2}(\theta
- \theta_n)^2 \mathcal{\ddot{V}}_n(\tilde{\theta}_n),\label{TAY}
\end{equation}
\indent~~~~ where $\tilde{\theta}_n$ is between $\theta $ and $ \theta_n.$\\
From the equality \ref{specifiedsequence1} combined with
\ref{residual} and by simple calculation, it is easy to check
that,
$$\mathcal{\ddot{V}}_n(\tilde{\theta}_n)=-\frac{1}{\sqrt{n}}\sum_{i=1}^{n}\ddot{M}_{f}(Y_i  - \tilde{\theta}_n Y_{i -1}) Y^2_{i-1}G(\textbf{Y}_i).$$
From the condition $(E.1)$, there exists some   positive constant
$K,$ such that, $$\mathcal{\ddot{V}}_n(\tilde{\theta}_n ) \leq K
\frac{1}{\sqrt{n}}\sum_{i=1}^{n} Y^2_{i-1}|G(\textbf{Y}_i)|.$$
Thus, we have the following, $$\frac{1}{2}(\theta - \theta_n)^2
\mathcal{\ddot{V}}_n(\tilde{\theta}_n)\leq \frac{K}{2}(\theta -
\theta_n)^2 \frac{1}{\sqrt{n}}\sum_{i=1}^{n}
Y^2_{i-1}|G(\textbf{Y}_i)|\leq\frac{K}{2\sqrt{n}}~\times[\sqrt{n}(\theta
- \theta_n)]^2 ~\times\frac{1}{n}\sum_{i=1}^{n}
Y^2_{i-1}|G(\textbf{Y}_i)|. $$
 By   Cauchy Schwartz inequality, it follows,\begin{equation} \mathbb{E}(Y^2_{i-1}|G(\textbf{Y}_i)|) \leq
 \sqrt{\mathbb{E}(\{Y^2_{i-1}\}^2)} \times \sqrt{\mathbb{E}(|G(\textbf{Y}_i)|^2)}.\label{cs}\end{equation}
Under $(C.2)$ and $(C.3)$ , we obtain,
$~~\mathbb{E}(Y^2_{i-1}|G(\textbf{Y}_i)|)< \infty.$
 \noindent Using the ergodicity and the stationarity of the model
\ref{FIRSTAR1MODEL}, it follows that $\frac{1}{n}\sum_{i=1}^{n}
Y^2_{i-1}|G(\textbf{Y}_i)|$ converge a.s. to some constant $k_1$
as $n \rightarrow \infty.$
 Since $[\sqrt{n}(\theta - \theta_n)]^2=O_p(1),$ then, it follows
 that,$\frac{1}{2}(\theta - \theta_n)^2 \mathcal{\ddot{V}}_n(\tilde{\theta}_n)=o_p(1).$ \\
Hence, the equality \ref{TAY}, will be rewritten,
\begin{equation}
\mathcal{V}_n(\theta) - \mathcal{V}_n(\theta_n) = \sqrt{n}(\theta
- \theta_n)\times \frac{\mathcal{\dot{V}}_n(\theta_n)}{\sqrt{n}}
+o_p(1).\label{TAYLOR1}
\end{equation}
\noindent This implies with the us of the Proposition
\ref{evaluation} the existence of  a random variable $ R_n$ which
converges in probability to $0$ and satisfying,
$$\sqrt{n}(\theta - \theta_n)= \sqrt{n}(\theta_{[1 +n^S]} - \theta_n) + R_n,$$
which enables us  to rewrite the equality (\ref{TAY}) as follows,
\begin{equation}
\mathcal{V}_n(\theta) - \mathcal{V}_n(\theta_n) =
\{\sqrt{n}(\theta_{[1 +n^S]} - \theta_n) + R_n\}\times
\frac{\mathcal{\dot{V}}_n(\theta_n)}{\sqrt{n}}\\=
\{\sqrt{n}(\theta_{[1 +n^S]} - \theta_n)\} \times
\frac{\mathcal{\dot{V}}_n(\theta_n)}{\sqrt{n}}~ +~ R_n \times
\frac{\mathcal{\dot{V}}_n(\theta_n)}{\sqrt{n}} +o_p(1).
\end{equation}
This gives under condition $(C.1),$ $$R_n \times
\frac{\mathcal{\dot{V}}_n(\theta_n)}{\sqrt{n}}=o_{P}(1).$$ In a
last, we deduce that,
\begin{equation}
\mathcal{V}_n(\theta)  =  \mathcal{V}_n(\theta_n) ~+~ (\theta_{[1
+n^S]} - \theta_n) {\mathcal{\dot{V}}_n(\theta_n)}) +
o_{P}(1)=\mathcal{W^S}_n(\theta_n) + o_{P}(1).
\end{equation}

 \item \noindent We have the following decomposition, $$\sqrt{n}
(\theta_{[1 +n^S]} -\theta_n )= \sqrt{n} (\theta_{[1 +n^S]}
-\theta ) + \sqrt{n} (\theta - \theta_n).$$ Obviously, the
quantities $\sqrt{n} (\theta_{[1 +n^S]} -\theta )$ and $\sqrt{n}
(\theta - \theta_n)$ are bounded, therefore, it follows that,
\begin{equation}\sqrt{n}
(\theta_{[1 +n^S]} -\theta_n )= O_{P}(1).\label{O}\end{equation}
Remark that $${D^S}_n=(\theta_{[1 +n^S]} -\theta_n
)\mathcal{\dot{V}}_n(\theta_n)=  \sqrt{n} (\theta_{[1 +n^S]}
-\theta_n ) \times
\frac{\mathcal{\dot{V}}_n(\theta_n)}{\sqrt{n}}.$$
 The combination of \ref{O} with the assumption $(C.1)$ implies that the random
variable ${D^S}_n=O_{P}(1).$\\
\noindent Since the  function $\theta\longrightarrow
\mathcal{V}_n(\theta)$ is derivable on $\mathring{\Theta}$, we
shall define the tangent space $\Gamma$ of the map $\mathcal{V}_n$
at $\theta_n$ by the following equation.\begin{equation} \Gamma:
\mathcal{V}_n(x) - \mathcal{V}_n(\theta_n)=
\mathcal{\dot{V}}_n(\theta_n)(x - \theta_n).\label{pert1}
\end{equation}
We use the same technical introduced in \cite{TL2011} [Subsection
2]. In other words, we search in the tangent space $\Gamma$ the
modified estimator $\bar{\theta}_n= \theta_n  + p,$ which
absorbing asymptotically the error ${D^S}_n $ and satisfying the
following,
\begin{equation}
\Gamma: \mathcal{V}_n(\bar{\theta}_n) - \mathcal{V}_n(\theta_n)=
{D^S}_n.\label{pert2}
\end{equation}
Therefore, from the combination of the equalities \ref{pert1} and
\ref{pert2}, we deduce that, $$p =\theta_{[1 +n^S]} -\theta_n
\mbox{~~and~~}\bar{\theta}_n= \theta_n  + p=\theta_{[1+n^S]}.$$
 Notice that, in this case, the modified estimator has an explicit
 form which the useful is easy  in practice. Recall that the assumptions $(C.0)$ and
$(C.1)$ were fixed in \cite{TL2011} for the existence and
consistency of this kind of estimators. Consequently, it follows
that, $\mathcal{V}_n(\theta_{[1 +n^S]}) ={D^S}_n +
\mathcal{V}_n(\theta_n)=\mathcal{W^S}_n(\theta_n).$\\ Making use
of the Proposition \ref{EQ1}, it results, that the sequences
 $\mathcal{V}_n(\theta_{[1 +n^S]})$ and $\mathcal{V}_n(\theta)$ are
equivalent with $o_{P}(1)$ close. Thus, the proof is complete.
\end{enumerate}
\noindent {\bf Proof of the Theorem \ref{optimaltest1}}
\noindent From \ref{lastequivalence1} and under $H_0$ the central
sequences $ \mathcal{V}_n(\theta)$  and $ \mathcal{V}_n(\theta_{[1
+n^S]})$ are equivalence with $o_{P}(1)$ close. The replacing of
the central sequence by the estimated central sequence  $
\mathcal{V}_n(\theta_{[1 +n^S]})$ has no effect. Under $(A.1)$ and
$(A.2)$ the local asymptotic normality of the logarithm ratio is
established, which implies the contiguity of the null and
alternative hypothesis (see for instance, \cite[Corrolary
4.3]{Droesbeke}). With the application of
 Le Cam third lemma's,  we obtain under  $H^{(n)}_1$
$\mathcal{V}_n \stackrel{\mathcal{D}}{\longrightarrow}
\mathcal{N}({\tau}^2,{\tau}^2).$ It follows from the convergence
in probability of  the estimate $\theta_{[1 +n^S]}$ to $\theta$ as
$n \longrightarrow+\infty$, and the application of the continuous
mapping theorem ( see, for instance \cite{W})
  on the continuous function $\tau : \cdot \longrightarrow
 \tau(\cdot)$  that asymptotically, the power of the test is no effected
 by replacing of  the unspecified parameter $\theta$ by it's estimator,
$\theta_{[1 +n^S]}$,  hence the optimality. By following the same
reasoning as \cite[Theorem 3]{hb}, we derive the asymptotic power
function $1 - \Phi(Z(\alpha)-{\tau}^2(\theta_{[1 +n^S]}))$.\\
\noindent {\bf Proof of the Proposition \ref{EQ3}}
\begin{enumerate}
\item At first, let be $N_f(x)=1 + x M_f(x)\label{N}.$ By simple
calculation, we shall prove that,
$$ \ddot{N}_f(x)=2\dot{M}_f(x) + x \ddot{M}_f(x).$$ From $(E.1)$,
in connection with the last equality, it result that, the function
$x\rightarrow
\ddot{N}_f(x)$ is bounded.\\
Also, the equality \ref{secondcentral} will be rewritten as
follows,\begin{eqnarray}
\mathcal{V}_{n}(\theta_n)&=&-\frac{1}{\sqrt{n}}\left\{\sum_{i=1}^{n}M_{f}(\epsilon_i)G(\textbf{Y}_i)
+\sum_{i=1}^{n}N_{f}(\epsilon_i)L(\textbf{Y}_i)\right\}.\label{secondcentral2}
 \end{eqnarray}
\noindent By simple calculation and with considering of the
equality \ref{secondcentral}, we obtain
$$\mathcal{\ddot{V}}_n(\tilde{\theta}_n)=-\frac{1}{\sqrt{n}}\sum_{i=1}^{n} Y^2_{i-1}G(\textbf{Y}_i)\ddot{M}_{f}(Y_i
- \tilde{\theta}_n Y_{i -1}) -\frac{1}{\sqrt{n}}\sum_{i=1}^{n}
Y^2_{i-1}L(\textbf{Y}_i)\ddot{N}_{f}(Y_i - \tilde{\theta}_n Y_{i
-1}).$$ From $(E.1),$ and since $x\rightarrow \ddot{N}_f(x)$ is
bounded,  there exists two positive constants $K_2$ and  $K_3$
such that,
 $$\mathcal{\ddot{V}}_n(\tilde{\theta}_n)\leq K_2 \times \big\{
\frac{1}{\sqrt{n}}\sum_{i=1}^{n}Y^2_{i-1}|G(\textbf{Y}_i)|\big\} +
K_3 \times
\big\{\frac{1}{\sqrt{n}}\sum_{i=1}^{n}Y^2_{i-1}|L(\textbf{Y}_i)|\big\}
.$$
 By  following the same reasoning as in the proof of the
Proposition \ref{EQ1}, we achieve the demonstration. \item The
reasoning is similar as in the Proposition \ref{EQ1}.
\end{enumerate}
\noindent {\bf Proof of the Theorem \ref{optimaltest2}}
\noindent The proof is analogous as in Theorem \ref{optimaltest1}.
\\\noindent {\bf Proof of the Proposition \ref{link}}
\noindent\begin{eqnarray}
\mathcal{V}_{n}(\theta)&=&-\frac{1}{\sqrt{n}}\left\{\sum_{i=1}^{n}M_{f}(\epsilon_i)G(\textbf{Y}_i)
+\sum_{i=1}^{n}(N_{f}(\epsilon_i)L(\textbf{Y}_i)\right\},\label{secondcentral11}
\end{eqnarray}
where $N_f$ is defined in Proposition \ref{EQ3}. By Taylor
expansion with order $2$ at $\theta_n$, we obtain\begin{eqnarray}
\mathcal{V}_n(\theta) = \mathcal{V}_n(\theta_n)
~+~\nabla\mathcal{V}_n(\theta_n)\cdot(\theta -\theta_n) ~+~
\frac{1}{2} (\theta
-\theta_n)^\prime\partial^2\mathcal{V}_n(\tilde{\theta}_n)(\theta
-\theta_n),\label{Taylor expa}
\end{eqnarray}
\indent where $\tilde{\theta}_n$ is between $\theta_n$ and
$\theta,~~$
${\tilde{\theta}_{n}}^\prime=\Big(\tilde{\theta}_{n,1},\dots,\tilde{\theta}_{n,m}),~~$
 $ \partial^2\mathcal{V}_n(\tilde{\theta}_n)=\left(%
\begin{array}{ccccc}
  D_{1,1}(\tilde{\theta}_n) & \cdot & \cdot & \cdot &D_{1,m}(\tilde{\theta}_n)  \\
  D_{2,1}(\tilde{\theta}_n) &  D_{2,2}(\tilde{\theta}_n) & \cdot & \cdot &D_{2,m}(\tilde{\theta}_n) \\
 \cdot & \cdot & \cdot & \cdot & \cdot \\
  \cdot & \cdot & \cdot & \cdot & \cdot \\
   D_{m,1}(\tilde{\theta}_n) &  D_{m,2}(\tilde{\theta}_n) & \cdot & \cdot &D_{m,m}(\tilde{\theta}_n) \\
\end{array}%
\right),$
 \noindent where, for $k,j=1\dots m$,~~
$$D_{k,k}(\tilde{\theta}_n)=\frac{
\partial^2\mathcal{V}_n(\tilde{\theta}_n)}{\partial^2\theta_k},
\mbox{~~and,~~}D_{k,j}(\tilde{\theta}_n)=\frac{\partial^2\mathcal{V}_n(\tilde{\theta}_n)}
 {\partial\theta_k \partial\theta_j}.$$
By simple calculation, it is easy to check that,
\begin{eqnarray}D_{k,k}(\tilde{\theta}_n)=-\frac{1}{\sqrt{n}}\sum_{i=1}^{n}\ddot{M}_{f}(Y_i
-  \sum_{j=1}^{m}\tilde{\theta}_{n,j}  Y_{i-j})
G(\textbf{Y}_i)Y^2_{i-k}
-\frac{1}{\sqrt{n}}\sum_{i=1}^{n}\ddot{N}_{f}(Y_i -
\sum_{j=1}^{m}\tilde{\theta}_{n,j}  Y_{i-j})
L(\textbf{Y}_i)Y^2_{i-k}.\nonumber\\\label{derivative1}\\
D_{k,j}(\tilde{\theta}_n)=-\frac{1}{\sqrt{n}}\sum_{i=1}^{n}\ddot{M}_{f}(Y_i
-  \sum_{l=1}^{m}\tilde{\theta}_{n,l}  Y_{i-l})
G(\textbf{Y}_i)Y_{i-k}Y_{i-j}
-\frac{1}{\sqrt{n}}\sum_{i=1}^{n}\ddot{N}_{f}(Y_i -
\sum_{l=1}^{m}\tilde{\theta}_{n,l}  Y_{i-l})
L(\textbf{Y}_i)Y_{i-k}Y_{i-j}.\nonumber\\\label{derivative2}
\end{eqnarray}
It remains to prove that the quantity $\frac{1}{2} (\theta
-\theta_n)^\prime\partial^2 \mathcal{V}_n(\tilde{\theta}_n)(\theta
-\theta_n)=o_p(1).$  By simple calculation, we obtain,
\begin{eqnarray}
\frac{1}{2} (\theta -\theta_n)^\prime\partial^2
\mathcal{V}_n((\tilde{\theta}_n)(\theta -\theta_n)=
\sum_{k=1}^{m}P_{k,n} +\sum_{j=1}^{m} Q_{j,n},  \mbox{~~where,}\nonumber\\
P_{k,n}=\frac{1}{2}(\theta_k-\theta_{k,n})^2D_{k,k}(\tilde{\theta}_n)
\mbox{~~and,~~} Q_{j,n}= \frac{1}{2} (\theta_j
-\theta_{j,n})\sum_{k=1,k \neq j}^{m}
D_{k,j}(\tilde{\theta}_n)(\theta_k
-\theta_{k,n}).\label{partielterm}\end{eqnarray} This, implies in
connection with \ref{derivative1} and \ref{derivative2},
that\begin{eqnarray}\frac{1}{2} (\theta_k -
\theta_{k,n})^2D_{k,k}(\tilde{\theta}_n)= \frac{1}{2} (\theta_k -
\theta_{k,n})^2 \times\big[
-\frac{1}{\sqrt{n}}\sum_{i=1}^{n}\{\ddot{M}_{f}(Y_i -
\sum_{j=1}^{m}\tilde{\theta}_{n,j})G(\textbf{Y}_i) +
\ddot{N}_{f}(Y_i - \sum_{j=1}^{m}\tilde{\theta}_{n,j}  Y_{i-j})
L(\textbf{Y}_i)\} \times Y^2_{i -k}\big].\nonumber\\
\label{firstinequality} \\ \frac{1}{2} (\theta_j
-\theta_{j,n})\sum_{k=1,k \neq j}^{m}
D_{k,j}(\tilde{\theta}_n)(\theta_k -\theta_{k,n})= \frac{1}{2}
(\theta_j -\theta_{j,n})\sum_{k=1,k \neq j}^{m} (\theta_k
-\theta_{k,n})\big\{
-\frac{1}{\sqrt{n}}\sum_{i=1}^{n}\ddot{M}_{f}(Y_i -
\sum_{l=1}^{m}\tilde{\theta}_{n,l}  Y_{i-l})
G(\textbf{Y}_i)Y_{i-k}Y_{i-j}\nonumber\\
-\frac{1}{\sqrt{n}}\sum_{i=1}^{n}\ddot{N}_{f}(Y_i -
\sum_{l=1}^{m}\tilde{\theta}_{n,l}  Y_{i-l})
L(\textbf{Y}_i)Y_{i-k}Y_{i-j}\big\}.\nonumber
\\\label{secondinequality}
\end{eqnarray}
From (E.1) and because $\ddot{N}_f$ is bounded, there exists two
positive constants $k_1$ and $k_2$ such that,
\begin{eqnarray*}\frac{1}{2} (\theta_k -
\theta_{k,n})^2|D_{k,k}(\tilde{\theta}_n)|\leq \frac{1}{2}
(\theta_k - \theta_{k,n})^2 \times
\big\{\frac{k_1}{\sqrt{n}}\sum_{i=1}^{n}
|G(\textbf{Y}_i)|Y^2_{i-k} ~+~\frac{k_2 }{\sqrt{n}}\sum_{i=1}^{n}
|L(\textbf{Y}_i)|Y^2_{i-k}\big\},\\ \leq \frac{1}{2}
[\sqrt{n}(\theta_k - \theta_{k,n})]^2 \times
\big\{\frac{k_1}{n}\sum_{i=1}^{n} |G(\textbf{Y}_i)|Y^2_{i-k}
~+~\frac{k_2 }{n}\sum_{i=1}^{n} |L(\textbf{Y}_i)|Y^2_{i-k}
\big\}\times \frac{1}{\sqrt{n}}.
\end{eqnarray*}
As proved in \ref{cs} and under  $(B.1)$, $(B.4)$, $(C.2)$ and
$(C.3)$ , we obtain,
$$\mathbb{E}(Y^2_{i-1}|G(\textbf{Y}_i)|)< \infty \mbox{~~and~~}\mathbb{E}(Y^2_{i-1}|L(\textbf{Y}_i)|)< \infty.$$
 From the ergodicity and stationary of model
\ref{modelprincipal2}, it follows that
$$\big\{\frac{k_1}{n}\sum_{i=1}^{n} |G(\textbf{Y}_i)|Y^2_{i-k}
~+~\frac{k_2 }{n}\sum_{i=1}^{n} |L(\textbf{Y}_i)|Y^2_{i-k}
\big\}\stackrel{a.s.}{\longrightarrow}
c_3=\mathbb{E}\big[k_1|G(\textbf{Y}_0)|Y^2_{-k} +
(k_2)|L(\textbf{Y}_0)|Y^2_{-k}\big]
\mbox{~as~}n\longrightarrow\infty.$$ Since the quantity
$\sqrt{n}(\theta_k - \theta_{k,n})$ is bounded, we conclude
that,\begin{eqnarray*}\frac{1}{2} (\theta_k - \theta_{k,n})^2
D_{k,k}(\tilde{\theta}_n)= P_{k,n}= o_p(1).\end{eqnarray*} Hence,

\begin{eqnarray} \sum_{k=1}^{m}P_{k,n}= o_p(1).\label{firstpetito}\end{eqnarray}
It remains to evaluate
the  quantity $Q_{j,n}.$\\
At first, we can see that, for all $1\leq j,k\leq m, j\neq k,$ we
have,\begin{eqnarray*}\frac{1}{2} (\theta_j -\theta_{j,n})
D_{k,j}(\tilde{\theta}_n)(\theta_k -\theta_{k,n})\leq\frac{1}{2}
\big|(\theta_j -\theta_{j,n})(\theta_k -\theta_{k,n})\big|\big\{
\frac{1}{\sqrt{n}}\big|\sum_{i=1}^{n}\ddot{M}_{f}(Y_i -
\sum_{l=1}^{m}\tilde{\theta}_{n,l}  Y_{i-l})
G(\textbf{Y}_i)Y_{i-k}Y_{i-j}\nonumber\\
+\frac{1}{\sqrt{n}}\sum_{i=1}^{n}\ddot{N}_{f}(Y_i -
\sum_{l=1}^{m}\tilde{\theta}_{n,l}  Y_{i-l})
L(\textbf{Y}_i)Y_{i-k}Y_{i-j}\big|\big\}.\nonumber\\\label{secondinequality1}
\end{eqnarray*}
 Since $|Y_{i-k} Y_{i-j}| \leq \frac{1}{2} \big[{Y_{i-k}}^2 + {Y_{i-j}}^2\big]$, and by
following the same previous reasoning as in the proof of
\ref{firstpetito}, we obtain,
$$\frac{1}{2} (\theta_j -\theta_{j,n})
D_{k,j}(\tilde{\theta}_n)(\theta_k -\theta_{k,n})=o_p(1).$$ Thus,
implies, $$ \frac{1}{2} (\theta_j -\theta_{j,n})\sum_{k=1,k \neq
j}^{m} D_{k,j}(\tilde{\theta}_n)(\theta_k
-\theta_{k,n})=Q_{j,n}=o_p(1),
  \mbox{~~ and,~~}  \sum_{j=1}^{m} Q_{j,n}=o_p(1).$$
The combination of  the last  previous equality with
\ref{firstpetito} implies that,
    $$ \frac{1}{2} (\theta
-\theta_n)^\prime\partial^2 \mathcal{V}_n(\tilde{\theta}_n)(\theta
-\theta_n)=o_p(1).$$ The subsisting  of the latter quantity into
 \ref{Taylor expa} achieve the proof.
 \subsection*{Proof of the Proposition \ref{delta method and confidence
 intervals}}
\noindent Note that Delta method enables us to deduce the limit
law of $\phi(\theta_n) - \phi(\theta)$ from that of $\theta_n  -
\theta,$ where $\phi$ is a function defined on a neighborhood of
$\theta$ and differentiable at $\theta.$ For more detail refer to
\cite[Chapter $3$]{W}. For each $j\in\{1,\dots, m\}$, we define
the function $\phi_j: {\mathbb{R}}^m \longrightarrow \mathbb{R},$
by $\phi_j(x_1,\dots,x_m)=x_j. $ \\
\noindent Making use of \cite[Theorem $3.1$]{W}, we deduce from
$(R.1)$ that,
 \begin{eqnarray*} \sqrt{n}(\phi_j(\theta_n) -
\phi_j(\theta)\stackrel{\mathcal{D}}{\longrightarrow}
\mathcal{N}(\phi_j(\theta)\cdot 0,\dot{\phi_j}(\theta)\Sigma
{\dot{\phi_j}(\theta)}^\prime),{~~where~~}\label{regularityimage}\\
\dot{\phi_j}(\theta)=\big[\frac {\partial\phi_j(\theta)}{\partial
x_1},\dots, \frac {\partial\phi_j(\theta)}{\partial x_m}\big],
\mbox{~~such that for~}i\neq j~~, \frac
{\partial\phi_j(\theta)}{\partial x_i}=0, \mbox{~and~} \frac
{\partial\phi_j(\theta)}{\partial x_j}=1.
\end{eqnarray*}
Thus, implies that, $ \sqrt{n}(\theta_{n,j}
-\theta_j)\stackrel{\mathcal{D}}{\longrightarrow}
\mathcal{N}(0,\Sigma_{j,j}).$ Then, we obtain, $ \frac
{\sqrt{n}(\theta_{n,j}
-\theta_j)}{\sqrt{\tilde{\Sigma}_{j,j}}}\stackrel{\mathcal{D}}{\longrightarrow}T(n
-m),$ where $T(n -m)$ is a Student distribution with the number of
degrees of freedom equal to $n- m$.
 \\\noindent {\bf Proof of the Lemma \ref{evaluationm }}
 \noindent For each j in $\{1,\dots,m\}$, we follow a same reasoning as in Lemma
 \ref{evaluation}.\\
\noindent {\bf Proof of the Proposition \ref{EQ5}}
\begin{enumerate}
    \item\noindent From the equality \ref{linkgeneralcase}, we
    have,
\begin{eqnarray}\mathcal{V}_n(\theta) - \mathcal{V}_n(\theta_n)=\nabla\mathcal{V}_{n}(\theta_n)\cdot(\theta  -\theta_N) +
\nabla\mathcal{V}_{n}(\theta_n)\cdot(\theta_N  -\theta_n)~+~
o_p(1).\label{equim}\end{eqnarray} Remark that,
$$\nabla\mathcal{V}_{n}(\theta_n)\cdot(\theta
-\theta_N)=\sum_{i=1}^{m}\frac{\partial
\mathcal{V}_{n}(\theta_n)}{\partial \theta_i}\times(\theta_i
-\theta_{N,i})=\sum_{i=1}^{m}\frac{\frac{\partial
\mathcal{V}_{n}(\theta_n)}{\partial
\theta_i}}{\sqrt{n}}\times\sqrt{n}(\theta_i -\theta_{N,i}). $$
The combination of $(C.1)$ with Lemma \ref{evaluationm } gives,
\begin{eqnarray}\frac{\partial \mathcal{V}_{n}(\theta_n)}{\partial
\theta_i}\times(\theta_i
-\theta_{N,i})=o_p(1).\label{paeteq}\end{eqnarray} Using
$\ref{paeteq}$, we obtain,
 $$\nabla\mathcal{V}_{n}(\theta_n)\cdot(\theta
-\theta_N)=o_p(1).$$ The subsisting of the latter into \ref{equim}
gives,\begin{eqnarray}\mathcal{V}_n(\theta -
\mathcal{V}_n(\theta_n)=\nabla\mathcal{V}_{n}(\theta_n)\cdot(\theta_N
-\theta_n)~+~ o_p(1).\label{equimfinal}\end{eqnarray}.
 \item Consider again the tangent space $\Gamma$  of the map
 $\mathcal{V}_n$ at $\theta_n,$ then it follows that,
\begin{eqnarray}\Gamma :\mathcal{V}_n(x)
 -\mathcal{V}_n(\theta_n)=\nabla\mathcal{V}_{n}(\theta_n)\cdot(x-\theta_n), \mbox{~~~where~} x \in \mathbb{R}^m\label{tangent}.\end{eqnarray}
\end{enumerate}
We search an estimator ${\bar{\theta}^j}_n $, with
$({\bar{\theta}^j}_n; \mathcal{V}_n({\bar{\theta}^j}_n)) \in
\Gamma$ and such that, ${\bar{\theta}^j}_{n,k}=\theta_{n,k}, k\neq
j,\mbox{~~and~~}{\bar{\theta}^j}_{n,j}=\theta_{n,j} +\rho_j.$\\
${\bar{\theta}^j}_n$ is obtained from the estimator $\theta_n$
with the perturbation of $j$-th component of $\theta_n$ and
satisfying,\begin{eqnarray}\mathcal{V}_n({\bar{\theta}^j}_{n,k})-\mathcal{V}_n(\theta_n)=\nabla\mathcal{V}_{n}(\theta_n)\cdot(\theta_N-\theta_n).\label{contrainte}
\end{eqnarray}
 With the subsisting  ${\bar{\theta}^j}_n $ into \ref{tangent} in
 connection with \ref{contrainte}, we obtain,
  $$\rho_j \times \frac{\partial
  \mathcal{V}_{n}(\theta_n)}{\partial \rho_j }=\nabla\mathcal{V}_{n}(\theta_n)\cdot(\theta_N-\theta_n).$$
  Hence,
         $$ \rho_j =\frac{\nabla\mathcal{V}_{n}(\theta_n)\cdot(\theta_N-\theta_n)}{\frac{\partial
  \mathcal{V}_{n}(\theta_n)}{\partial \rho_j }}.$$
  Then, we deduce,
  $${\bar{\theta}^j}_n  =\big(\theta_{n,1};\dots;\theta_{n,j-1}; \theta_{n,j} + \rho_j ;\theta_{n,j+1};\dots;\theta_{n,m} \big)^\prime.$$

\noindent {\bf Proof of the Theorem \ref{optimaltest3}}
The proof is similar as in Theorem \ref{optimaltest1}.

\end{document}